\documentclass[twocolumn,showpacs,amsmath,amssymb,pra,superscriptaddress]{revtex4}

\usepackage{graphicx}% Include figure files
\usepackage{dcolumn}% Align table columns on decimal point
\usepackage{bm}% bold math
\usepackage{subfigure}
\begin{document}

\title{{\it Ab initio} based description of the unusual temperature increase of\\
the electric field gradient at Ti sites in rutile TiO$_2$
}

\author{A. V. Nikolaev}

\affiliation{Skobeltsyn Institute of Nuclear Physics Lomonosov
Moscow State University, 119991 Moscow, Russia}

\affiliation{National Research Nuclear University MEPhI, 115409,
Kashirskoe shosse 31, Moscow, Russia}

%\affiliation{School of Electronics, Photonics and Molecular Physics, Moscow Institute of Physics and Technology, 141700, Dolgoprudny, Moscow %region, Russia}
\affiliation{Moscow Institute of Physics and Technology, 141700, Dolgoprudny, Moscow Region, Russia}

\author{N. M. Chtchelkatchev}
\affiliation{Vereshchagin Institute for High Pressure Physics, Russian Academy of Sciences, 108840 Troitsk, Moscow, Russia}
\affiliation{Moscow Institute of Physics and Technology, 141700, Dolgoprudny, Moscow Region, Russia}

\author{A. V. Bibikov}

\affiliation{Skobeltsyn Institute of Nuclear Physics Lomonosov
Moscow State University, 119991 Moscow, Russia}

\affiliation{National Research Nuclear University MEPhI, 115409,
Kashirskoe shosse 31, Moscow, Russia}

\author{D. A. Salamatin}
\affiliation{Vereshchagin Institute for High Pressure Physics, RAS, 142190 Troitsk, Moscow, Russia}
\affiliation{Dzelepov Laboratory of Nuclear Problems, Joint Institute for Nuclear Research, 141980 Dubna, Russia}

\author{A. V. Tsvyashchenko}
\affiliation{Vereshchagin Institute for High Pressure Physics, RAS, 142190 Troitsk, Moscow, Russia}

%\affiliation{National Research Nuclear University MEPhI, 115409,
%Kashirskoe shosse 31, Moscow, Russia}

%\date{\today}

%--------------- ABSTRACT ---------------
\begin{abstract}
Combining a precise {\it ab initio} electron band structure calculation of the TiO$_2$ rutile
structure with the temperature evolution of the Ti mean-square displacements,
we reproduce a puzzling temperature increase of the electric field gradient at Ti sites in TiO$_2$,
observed experimentally.
Our method employs a procedure of averaging two quadrupole electron density components ($L = 2$) inside a sphere
vibrating with the Ti nucleus at its center, where the key factor introducing the temperature dependence
is the square root of the Debye-Waller factor.
Although the Debye-Waller factor always reduces the corresponding Fourier component,
in TiO$_2$ due to the interplay between terms of opposite signs, it
results in a net increase of the whole sum with temperature, leading to the growth of the electric field gradient.
Quantitatively, we find that the increase of electric field gradient is only half of the experimental value,
which we ascribe to anharmonic effects or a strong oxygen position influence.
In addition, our method reproduces the unusual temperature dependence of the asymmetry parameter $\eta$,
which first decreases with temperature, goes to zero and then increases.
\end{abstract}


\maketitle

%%%%%%%%%%%%%%%%%%%%%%%%%%%%%%%%%%%%%%%%%%%%%%%%%%%%%%%%%%%%%%%%%%%%
\section{Introduction}
\label{sec:int}

Nuclear spectroscopic methods -- such as the time-differential perturbed angular correlation (TDPAC) technique
and the M{\"o}ssbauer spectroscopy -- are very sensitive experimental instruments probing the local charge distributions
of complex materials of interest over a wide temperature ($T$) range \cite{Nuc,PAC,PAC19}.
These methods measure the electric field gradient (EFG) at nuclear probes \cite{Kauf,Haa0}, reflecting minute changes
in their chemical environment and local bonding.
In the case of a phase transition, EFG (or $V_{zz}$) can change abruptly \cite{Tsv}, but even without phase transitions
the EFG demonstrates a smooth change with $T$ \cite{Kauf,Haa0}.
Usually, this is a monotonic decrease with increasing $T$, very often in the functional form of
$V_{zz} \propto (1 - B T^{3/2})$ \cite{Chris,Kauf} (where $B$ is a constant)
but in some instances unexpectedly an increase of $V_{zz}$ with $T$
takes place. In particular, a weak increase of $V_{zz}$ with $T$ has been observed for the $^{47}$Ti and $^{49}$Ti nuclear probes
in the rutile structure of titanium dioxide~\cite{Kaner}.

The nature of a strong temperature behavior of $V_{zz}$ has been addressed in a number of theoretical
and experimental research \cite{Kauf,Haa0,Nik}. In principle, one can consider it as an anharmonic effect.
Then, taking the experimental temperature dependence of lattice constants -- $a(T)$ and $c(T)$ for TiO$_2$ for example --
and performing a set of EFG calculations at distinct values of $a_i$ and $c_i$ corresponding to various temperatures $T_i$,
one can obtain the evolution of $V_{zz}$ as a function of $T$.
This is of course not true {\it ab initio} approach, but even worse -- the method in many cases does not work,
giving wrong dependence: a decrease instead of increase and vice versa.
This was realized already at the early stage of investigation albeit for a simple point charge model \cite{Kauf}.
The challenge is to explain the temperature evolution of $V_{zz}$ in the harmonic approximation, at least
its general tendency to increase or decrease with $T$.
To date there have been only a few attempts of such research \cite{Toru,Nik}.

A recent study \cite{Nik} has combined an {\it ab initio} band structure approach with
computing Debye-Waller factors (DWFs), $\exp(-W(K,T))$ \cite{Bru},
to reproduce correctly the experimental temperature evolution of EFG in pristine metallic zinc and cadmium \cite{Chris}.
This method is in accord with early idea of Jena \cite{Jen}, who used a simple empirical model
for the temperature dependence of DWFs and obtained a $(1 - B T^{3/2})$ decrease of EFG.
The model of Jena can only describe the reduction of $V_{zz}$ with $T$, which seems logical
taking into account the monotonic decrease of DWFs with temperature \cite{Bru}.
A fundamental question is --  can the consideration based on DWFs explain an increase of EFGs for example, in TiO$_2$?
As discussed in Ref.\ \cite{Nik}, in principle it is possible. However, it has been shown only that this type of
behaviour is not forbidden and can occur within a certain model dependence for the temperature evolution of mean square displacements.
In the present paper, we extend the considerations developed in Ref.\ \cite{Nik} to the rutile structure of TiO$_2$ and
demonstrate that the method is capable of predicting the increase of $V_{zz}$ with $T$.

\begin{figure}
\resizebox{0.40\textwidth}{!} {
\includegraphics{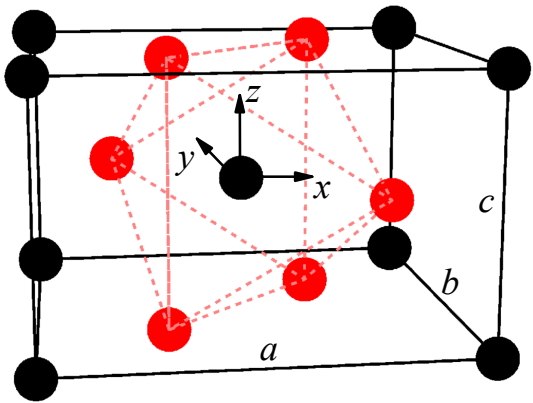}
}
% -----------> Figure Caption
\caption{
Unit cell of the rutile phase of TiO$_2$ and the Cartesian coordinate system.
(Titanium - black spheres, oxygen - red spheres.)
The distorted oxygen octahedron around the central Ti atom is shown by red dashed lines.
The lattice constants $a$, $b=a$, $c$ are indicated.
} \label{fig1}
\end{figure}

The rutile phase of TiO$_2$ is the thermodynamically stable form of the three known modifications
(rutile, anastase, brookite) of titanium dioxide \cite{Grant}. The Ti ions form the body centered tetragonal structure,
with each Ti ion surrounded by a distorted octahedron of six oxygen ions, Fig.\ \ref{fig1}.
In the rutile unit cell there are two different Ti sites distinguished by two alternating orientations
of the surrounded distorted oxygen octahedra.
These octahedra and two tensors of electric field gradients at two Ti sites are related by a rotation through $\pi$ about the $z-$axis.
Below in considering EFG tensor we refer to the central Ti site in Fig.\ \ref{fig1}.

The Ti-site EFGs have been measured with the native $^{47}$Ti and $^{49}$Ti nuclear probes
by means of the nuclear magnetic resonance (NMR) spectroscopy \cite{Kaner}.
These measurements performed on a single crystal over a large temperature range, indicate that
$V_{zz}$ increases from 2.11$\times$10$^{21}$~V/m$^2$ at 150~K to 2.86$\times$10$^{21}$~V/m$^2$ at 1420~K.
The second parameter $\eta$ -- the EFG asymmetry -- starts from 0.2 at 300~K, goes to zero around 970~K
and then increases to 0.13 at 1420~K \cite{Kaner,Butz}.
(The interpolated experimental plots for EFG and $\eta$ are shown in Fig.\ \ref{fig7} later.)
In addition, the TDPAC spectroscopy has been applied to TiO$_2$ to study EFG with
the impurity $^{111}$In$\rightarrow ^{111}$Cd \cite{Wenz,Adams,Erri02}, $^{181}$Hf$\rightarrow ^{181}$Ta \cite{Adams,Darr}
and $^{44}$Ti$\rightarrow ^{44}$Sc \cite{Ryu,Butz13} nuclear probes.
The $V_{zz}$ behavior of the $^{111}$Cd and $^{181}$Ta probes differs considerably from
that of Ti. In contrast to $^{181}$Ta TDPAC measurements \cite{Adams}
showing as $^{47,49}$Ti an increase with $T$, the
substitutional $^{111}$Cd probes result in a slow decrease of $V_{zz}$ with $T$ \cite{Adams}.
The room temperature values of $V_{zz}$ are 13.4$\times$10$^{21}$~V/m$^2$ for $^{181}$Ta
and 5.8$\times$10$^{21}$~V/m$^2$ $^{111}$Cd.
The asymmetry parameter $\eta$ with these impurities also changes differently. In $^{181}$Ta
$\eta$ increases slightly from 0.57 at 300 K to 0.58 at 1300~K, whereas
with $^{111}$Cd $\eta$ changes from 0.2 at 300~K goes to zero at 700~K and then increases to 0.2 at 1300~K.
%%%%%%%%%%%%%%%%%%%%%%%%%%%%%%%%%%%%%%%%%%%%%%%%%%%%%%%%%%%%%%%%%%
%
%------------------------------------------------------
%    FIGURE 1
%------------------------------------------------------
%
%%%%%%%%%%%%%%%%%%%%%%%%%%%%%%%%%%%%%%%%%%%%%%%%%%%%%%%%%%%%%%%%%
Although recent TDPAC measurements of $^{44}$Sc probes in rutile modification of TiO$_2$ found
that EFG increased with $T$ \cite{Butz13}, they also showed
a remarkable difference from the native $^{47}$Ti and $^{49}$Ti nuclear probes, detecting
essentially $T$-independent antiaxial EFG with $\eta=0.94$ close to unity \cite{Ryu,Butz13}.

The EFG at Ti sites has been a subject of band structure investigations \cite{Blaha,Erri02,Erri03,Erri05}.
In the early study of Blaha et al. \cite{Blaha} in pure TiO$_2$ it has been found that in addition to valence states,
there is a relatively large contribution to $V_{zz}$ from the semicore Ti 3$p$ electron states,
and that the value of $V_{zz}$ is very sensitive to details of the electron band structure.
In a series of first-principles calculations the problem of a substitutional Cd impurity
located at the Ti site has been addressed in Refs.\ \cite{Erri02,Erri03,Erri05}.
On substituting the Ti atom by a Cd impurity the {\it ab initio} calculation
predicts strong anisotropic relaxations of the nearest oxygen neighbors and a change of the
orientation of the largest EFG tensor component, $V_{33}$, from the (001) to the (110) direction \cite{Erri02}.
To calculate the temperature dependence of $V_{zz}$ at Cd the structural relaxation
has been performed for a set of experimental temperature dependent lattice constants $a$ and $c$ \cite{Erri05}.
This study however refers to a decrease of $V_{zz}$.

To the best of our knowledge no attempt was made to explain
the temperature increase of $V_{zz}$ in TiO$_2$.
This issue is the main goal of the present study.
In Sec. II we briefly describe the method.
Here we generally follow the considerations of Ref.\ \cite{Nik},
but apply the approach to the problem of TiO$_2$.
In Sec.\ III we present our results and discuss them.
The conclusions are summarized in Sec.~IV.

%%%%%%%%%%%%%%%%%%%%%%%%%%%%%%%%%%%%%%%%%%%%%%%%%%%%%%%%%%%%%%%%%%%%
\section{Theoretical Method}
\label{sec:method}

%%%%%%%%%%%%%%%%
\subsection{Electric Field Gradient in TiO$_2$}
\label{sub_EFG}

The EFG tensor $V_{ij}$ is defined as the second partial spatial derivatives of the electric part of the self-consistent-field potential $V(\vec{R})$
evaluated at the nuclear site, i.e.
\begin{eqnarray}
    V_{ij} = \frac{\partial^2 V}{\partial i \partial j} ,
\label{i1}
\end{eqnarray}
where $i = x, y, z$. $V_{ij}$ being a symmetric second rank tensor, can be further diagonalized
by transforming coordinates to the principal system of axes chosen such that $|V_{zz}| \ge |V_{yy}| \ge |V_{xx}|$.
(Since $V_{ij}$ is traceless, in the principal axis system the number of independent parameters for EFG is reduced to two.)
The principal component ($V_{zz}$) is called the electric field gradient, and the second independent parameter
is the asymmetry $\eta$ defined as $\eta = (V_{xx} - V_{yy})/V_{zz}$ ($0 \le \eta \le 1$).

%%%%%%%%%%%%%%%%%%%%%%%%%%%%%%%%%%%%%%%%%%%%%%%%%%%%%%%%%%%%%%%%%%
%
%------------------------------------------------------
%    FIGURE 2
%------------------------------------------------------
\begin{figure}
\resizebox{0.45\textwidth}{!} {
\includegraphics{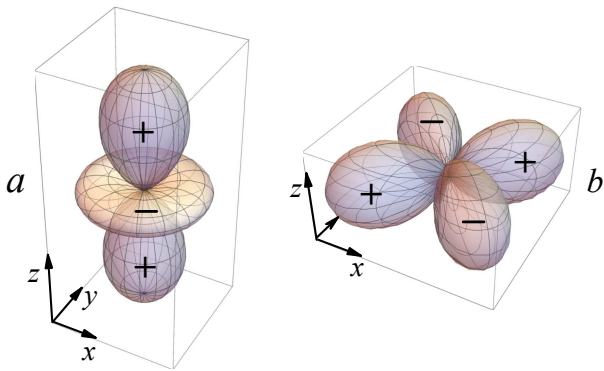}
}
% -----------> Figure Caption
\caption{
Two angular quadrupole functions, $S_{Q,1}$ ($a$) and $S_{Q,2}$ ($b$), Eqs.\ (\ref{e1a}) and (\ref{e1b}), used for the
expansion of the electron density and electric potential at the titanium site in TiO$_2$. The $x$, $y$ and $z$-axes
correspond to those shown in Fig.\ \ref{fig1}.
} \label{fig2}
\end{figure}
%
%%%%%%%%%%%%%%%%%%%%%%%%%%%%%%%%%%%%%%%%%%%%%%%%%%%%%%%%%%%%%%%%%
It can be shown (see e.g. Ref.\ \cite{Nik}) that the tensor of EFG $V_{ij}$ is closely related to the quadrupole components ($l=2$)
of the expansion of the electric part of the full potential $V(\vec{R})$ in terms of the angular functions $S_l^{\tau}(\theta,\phi)$,
\begin{eqnarray}
    V(r, \theta, \phi) = \sum_{l,\tau} V_{l}^{\tau}(r)\, S_{l}^{\tau}(\theta,\phi) .
\label{p1}
\end{eqnarray}
Here $S_{l}^{\tau}(\theta, \phi)$ are the angular functions adapted for the rutile point crystal symmetry,
$l$ is the multipole orbital index  and $\tau$ counts functions with the same $l$ (if there are few such functions).
The polar angles $(\theta, \phi)$ are specified by the vector $\vec{r}$ from the closest nuclear position.
$S_{l}^{\tau}(\theta, \phi)$ called symmetry adapted functions (SAFs) are linear combinations of real spherical harmonics
($Y_l^{m,c} \sim \cos m\phi$ and $Y_l^{m,s} \sim \sin m\phi$)
defined by the crystal site symmetry and tabulated in Ref. \cite{BC}. Eq.\ (\ref{p1}) can be viewed as a standard expansion in terms
of spherical harmonics, where each angular function has the full point crystal symmetry.
Expansion (\ref{p1}) is a standard procedure in the full potential version of modern electron band structure codes \cite{flapw}.

\begin{figure}
\resizebox{0.45\textwidth}{!} {
\includegraphics{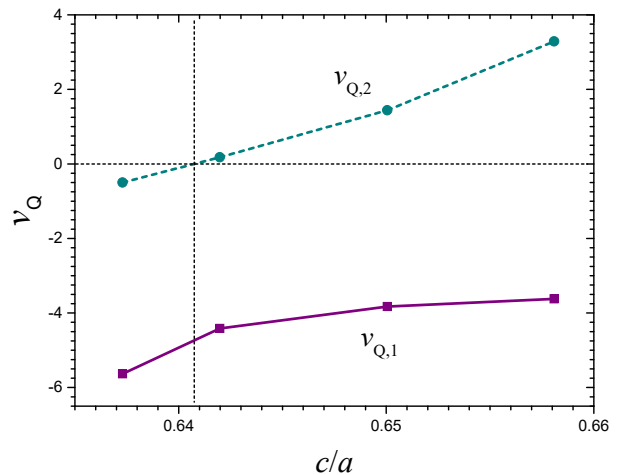}
}
% -----------> Figure Caption
\caption{
$v_{Q,1}$ and $v_{Q,2}$, Eq.\ (\ref{e3}),
versus the lattice constant ratio $c/a$ (with $u$ fixed) at $T=0$, in units (eV/a.u.$^2$).
For $c/a > 0.641$ $v_{Q,2}>0$ and $V_{11} > V_{22}$, for $c/a < 0.641$ $v_{Q,2}<0$ and $V_{11} < V_{22}$,
Eqs.\ (\ref{e4a}), (\ref{e4b}).
At $c/a \approx 0.641$ $\eta = 0$.
(Calculations with the Moscow-FLAPW code, details are given in Sec.\ \ref{sub_lapw}.)
} \label{fig3}
\end{figure}

At titanium site there are two quadrupole SAFs related to the tensor of EFG, Fig.\ \ref{fig2},
\begin{subequations}
\begin{eqnarray}
    S_{Q,1}(\theta,\phi) &=& Y_{l=2}^{0}(\theta,\phi)  \sim 3 z^2 - r^2,  \label{e1a} \\
    S_{Q,2}(\theta,\phi) &=& Y_{l=2}^{2,s}(\theta,\phi) \sim xy .  \label{e1b}
\end{eqnarray}
\end{subequations}
Here $Y_l^{m=0}$ and $Y_l^{m,s}$ are real spherical harmonics with $l=2$.
Thus, the quadrupole part of the electric potential, Eq.\ (\ref{p1}), is given by
\begin{eqnarray}
    V_Q(r,\, \theta,\phi) = V_{Q,1}(r)\, S_{Q,1}(\theta,\phi) + V_{Q,2}(r)\, S_{Q,2}(\theta,\phi) .
\label{e2}
\end{eqnarray}
Here $V_{Q,i}(r)$ ($i=1,2$) are two independent radial components corresponding to the angular parts $S_{Q,i}(\theta,\phi)$.
The tensor of EFG $V_{ij}$, Eq.\ (\ref{i1}), comes solely from the function $V_Q$, Eq.\ (\ref{e2}), the contribution from the other terms in Eq.\ (\ref{p1})
is zero.  As shown in Ref.\ \cite{Nik},
for the radial components $V_{Q,i}(r)$ at $r \rightarrow 0$ we obtain
\begin{eqnarray}
    V_{Q,1}(r) = v_{Q,1}\, r^2, \quad  V_{Q,2}(r) = v_{Q,2}\, r^2,
\label{e3}
\end{eqnarray}
where $v_{Q,1}$ and $v_{Q,2}$ are potential constants. The calculated dependencies of $v_{Q,1}$ and $v_{Q,2}$ on the
lattice constant ratio $c/a$ are shown in Fig.\ \ref{fig3}.
%%%%%%%%%%%%%%%%%%%%%%%%%%%%%%%%%%%%%%%%%%%%%%%%%%%%%%%%%%%%%%%%%%
%
%------------------------------------------------------
%    FIGURE 3
%------------------------------------------------------
%
%%%%%%%%%%%%%%%%%%%%%%%%%%%%%%%%%%%%%%%%%%%%%%%%%%%%%%%%%%%%%%%%%
From Eqs.\ (\ref{e2}) and (\ref{e3}) one can obtain the diagonal components of the EFG tensor,
\begin{subequations}
\begin{eqnarray}
    V_{11} &=& -v_{Q,1}\, \frac{1}{2} \sqrt{\frac{5}{\pi}} + v_{Q,2}\, \frac{1}{2} \sqrt{\frac{15}{\pi}} ,  \label{e4a} \\
    V_{22} &=& -v_{Q,1}\, \frac{1}{2} \sqrt{\frac{5}{\pi}} - v_{Q,2}\, \frac{1}{2} \sqrt{\frac{15}{\pi}} ,  \label{e4b} \\
    V_{33} &=& v_{Q,1}\, \sqrt{\frac{5}{\pi}} .  \label{e4c}
\end{eqnarray}
\end{subequations}
Here the diagonal component $V_{11}$ is oriented along the $(1,1,0)$ crystal lattice direction, $V_{22}$ along the $(-1,1,0)$ direction
and $V_{33}$ -- along the $(0,0,1)$ direction.

\begin{figure}
\resizebox{0.45\textwidth}{!} {
\includegraphics{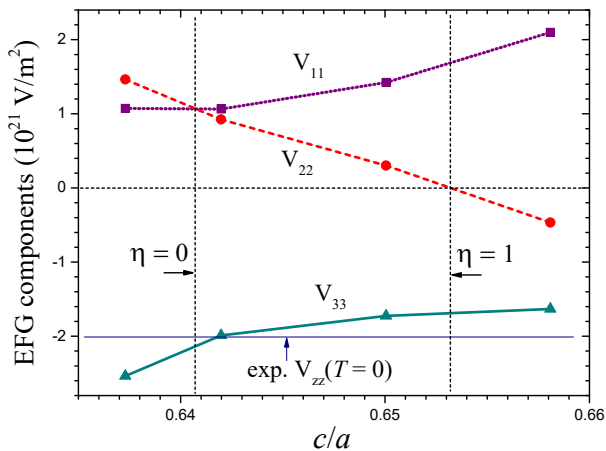}
}
% -----------> Figure Caption
\caption{
Diagonal components $V_{jj}$ ($j=1,2,3$) of the tensor of EFG, Eq.\ (\ref{e4a})--(\ref{e4c}),
versus the lattice constant ratio $c/a$ at $T=0$ (with $u$ fixed).
The electric field gradient $V_{zz}=V_{33}$ for $c/a < 0.653$, and $V_{zz}=V_{11}$ for $c/a > 0.653$.
At $c/a \approx 0.641$ $\eta = 0$, at $c/a \approx 0.653$ $\eta = 1$.
For comparison the $T=0$ value of $V_{zz}$ obtained by the extrapolation of experimental data \cite{Kaner} is also shown.
(Calculations with the Moscow-FLAPW code, details are given in Sec.\ \ref{sub_lapw}.)
} \label{fig4}
\end{figure}

The calculated dependence of the components $V_{jj}$ on the lattice constant ratio $c/a$ with the fixed $u=0.3041$ is depicted in Fig.\ \ref{fig4}.
%%%%%%%%%%%%%%%%%%%%%%%%%%%%%%%%%%%%%%%%%%%%%%%%%%%%%%%%%%%%%%%%%%
%
%------------------------------------------------------
%    FIGURE 4
%------------------------------------------------------
%
%%%%%%%%%%%%%%%%%%%%%%%%%%%%%%%%%%%%%%%%%%%%%%%%%%%%%%%%%%%%%%%%%
From Figs.\ \ref{fig3} and \ref{fig4} (see also Sec.\ \ref{sec:res}) it follows that in TiO$_2$ where $c/a \approx 0.640-0.644$,
$v_{Q,1} < 0$ and $V_{zz}=V_{33} < 0$ and hence $V_{11} > 0$, $V_{22}>0$.
That is, the EFG value is associated with the $V_{33}$ component
along the $(0,0,1)$ crystal lattice direction.
For $c/a < 0.641$ $v_{Q,2}<0$, $V_{11} < V_{22}$, and $V_{xx}=V_{11}$, $V_{yy}=V_{22}$, $\eta = (V_{11} - V_{22})/V_{33}$.
For $c/a > 0.641$ $v_{Q,2}>0$ and $\eta = (V_{22} - V_{11})/V_{33}$, Fig.\ \ref{fig3}.

Interestingly, at $c/a > 0.653$ the largest diagonal component becomes $V_{11}$ (i.e. $V_{zz}=V_{11}$), Fig.\ \ref{fig4}.
The change of $V_{zz}$ orientation from (001) to (110) directions is experimentally observed for Cd probes in Ti sites in TiO$_2$ \cite{Erri02,Erri03},
where it comes from the local anisotropic atomic relaxation introduced by the Cd impurity.
In pure TiO$_2$ the situation is different and the change of EFG is caused by the potential increase of $c/a$ modifying
the nearest neighbors' Ti-O relative distances (those from the apical
oxygens and those of the basal plane) and thus producing a similar effect.

Therefore, there are two critical points in Fig.\ \ref{fig4}.
At $c/a=0.641$ $\eta=0$ and the tensor of EFG is axial, whereas at $c/a=0.653$ $\eta=1$ and the tensor of EFG
is antiaxial \cite{Ryu}.

%%%%%%%%%%%%%%%%
\subsection{Temperature dependence of EFG}
\label{sub_T}

A general model for theoretical calculations of temperature dependence of EFG is presented in Ref.\ \cite{Nik}.
Here we briefly outline it applying to TiO$_2$ and working within the linear augmented plane wave (LAPW) method \cite{flapw}.

The electron density around a nucleus being strongly coupled to it by the Coulomb interaction
follows adiabatically the vibrating ion, while the electron density in the interstitial region
interacting with several neighboring ions is less affected by the motion of a single atom.
The dichotomy is resolved by considering vibrating spheres
[muffin-tin or MT-spheres] in the crystal with the electron density
virtually unchanged in the interstitial region (between the spheres).
That is, we assume that the Fourier expansion coefficients $\rho(\vec{K})$ of the electron density
in the interstitial density remain constant,
and, following the procedure of LAPW method, calculate multipole density components
on the vibrating MT spheres, which serve as the boundary surface conditions. Inside the MT-spheres the electron density
is expanded in terms of spherical harmonics.
As shown in Ref.\ \cite{Nik}
at a finite and even zero temperature $T$ the average quantities $\langle \rho(\vec{K}) \rangle$ on vibrating MT-spheres are effectively reduced.
In the language of the LAPW method this implies a modification of boundary conditions for the solutions inside the MT-spheres \cite{flapw}.
In particular, this procedure affects the average value of quadrupole components of electron density, $\langle \rho_{Q} \rangle$
associated with the multipole index $l=2$.
The change of $\langle \rho_{Q} \rangle$ on the sphere surface (at $r = R_{MT}$) translates inside it (i.e. at $r \le R_{MT}$) and
finally leads to a change of EFG at $r \rightarrow 0$.

Unlike Ref.\ \cite{Nik}, the point symmetry of the Ti site in TiO$_2$ supports two angular quadrupole functions -- $S_{Q,i}$, $i=1,2$, Eqs.\ (\ref{e1a}) and (\ref{e1b}) --
in the multipole expansions of the potential and electron desnity.
Extending the considerations of Ref.\ \cite{Nik}, we then write for
the average values of two quadrupole components of electron density on the Ti MT-sphere,
\begin{eqnarray}
    \langle \rho_{Q,i} \rangle = -4 \pi \sum_{\vec{K}} j_2(K R_{MT}) S_{Q,i}(\hat{K})\, e^{-W(\vec{K},T)}\, \rho(\vec{K}) .  \label{m1}
\end{eqnarray}
Here the summation is over the reciprocal vectors $\vec{K}$, $\hat{K}$ specifies the direction of $\vec{K}$, i.e. $\hat{K} \equiv (\theta_{\vec{K}},\phi_{\vec{K}})$,
$j_2$ are spherical Bessel functions,
$S_{Q,1}$ and $S_{Q,2}$ are the quadrupole SAFs given by Eqs.\ (\ref{e1a}) and (\ref{e1b}), and, finally,
\begin{eqnarray}
    W(\vec{K},T) = \frac{1}{2} \langle (\vec{K} \cdot \vec{u})^2 \rangle ,
\label{m2}
\end{eqnarray}
where $\vec{u}$ is the vector of the MT-sphere displacement.
In Eq.\ (\ref{m1}), the effective reductions of $\langle \rho(\vec{K}) \rangle$ is described by the temperature function $\exp(-W)$,
which is the square root of the Debye-Waller factor (SRDWF). (The conventional Debye-Waller factor is $\exp(-2W)$, Ref.\ \cite{Bru}.)
Earlier, SRDWF has been used by Kasowski in explaining the temperature-dependent Knight shift in cadmium \cite{Kas1,Kas2}.

As argued in Ref.\ \cite{Nik} in a first approximation one can use an $r$-independent change (often reduction)
of the quadrupole electron density component $\rho_{Q,i}(r)$ inside the Ti MT-sphere ($r \le R_{MT}$),
\begin{eqnarray}
   \frac{\langle \rho_{Q,i}(r,T) \rangle}{\rho_{Q,i}(r,\,u=0)} = \frac{\langle \rho_{Q,i}(R_{MT},T) \rangle}{\rho_{Q,i}(R_{MT},\,u=0)} = R_i(T) .
  \label{m3}
\end{eqnarray}
Here $\langle \rho_{Q,i}(r,T) \rangle$ is the $i$-th averaged quadrupole value inside the sphere ($i=1,2$), while
$\rho_{Q,i}(r,\,u=0)$ is the quadrupole component in the absence of displacements ($u=0$).
Note that $\langle \rho_{Q,i}(R_{MT},T) \rangle$ on the surface of Ti MT-sphere can be calculated by means of Eq.\ (\ref{m1}).
Combining Eqs.\ (\ref{m1}) and (\ref{m3}), we obtain $R_1$ and $R_2$ as functions of $T$. At $T=0$, $R_i(T) \approx 1$.
(The deviations of $R_i(T=0)$ from one are due to the zero point vibrations.)

The temperature changes of $\rho_{Q,i}$ inside the Ti MT-spheres described by Eq.\ (\ref{m3}) lead to the proportional changes of
the quadrupole potential components, i.e. $V_{Q,i}(r,T)=R_i(T)V_{Q,i}(r)$ \cite{Nik}. Eq.\ (\ref{e2}) then reads as
\begin{eqnarray}
    V_Q(r,\Omega,\, T) = V_{Q,1}(r,T)\, S_{Q,1}(\Omega) + V_{Q,2}(r,T)\, S_{Q,2}(\Omega) , \nonumber \\
\label{m3a}
\end{eqnarray}
where $\Omega=(\theta,\phi)$.
At $r \rightarrow 0$ we obtain $V_{Q,i}(r,T)=v_{Q,i}(T)\, r^2$ [compare with Eq.\ (\ref{e3})] where we have
introduced temperature dependent functions
\begin{eqnarray}
    v_{Q,1}(T) = v_{Q,1}\, R_1(T), \quad  v_{Q,2}(T) = v_{Q,2}\, R_2(T).
\label{m3b}
\end{eqnarray}
Then the temperature dependence of three components of EFG, Eqs.\ (\ref{e4a})--(\ref{e4c}) is excplicitly expressed as
\begin{subequations}
\begin{eqnarray}
    V_{11}(T) &=& -R_1(T)\, \frac{v_{Q,1}}{2} \sqrt{\frac{5}{\pi}} + R_2(T)\, \frac{v_{Q,2}}{2} \sqrt{\frac{15}{\pi}} , \quad \quad \label{m4a} \\
    V_{22}(T) &=& -R_1(T)\, \frac{v_{Q,1}}{2} \sqrt{\frac{5}{\pi}} - R_2(T)\, \frac{v_{Q,2}}{2} \sqrt{\frac{15}{\pi}} ,  \label{m4b} \\
    V_{33}(T) &=& R_1(T)\, v_{Q,1}\, \sqrt{\frac{5}{\pi}} .  \label{m4c}
\end{eqnarray}
\end{subequations}
As long as $V_{33}$ remains the largest diagonal component (i.e. $V_{zz}=V_{33}$) we arrive at
\begin{eqnarray}
   V_{zz}(T) \approx R_1(T) V_{zz} ,
  \label{m5}
\end{eqnarray}
where $V_{zz}$ is the $T=0$ value of EFG. On the other hand, the asymmetry parameter $\eta(T)$ depends on both $v_{Q,1}(T)$ and $v_{Q,2}(T)$.

In the following we will work in this approximation, and therefore, the only quantities that we have to calculate are two
ratios of quadrupole components on the surface of Ti MT-sphere $R_i(T)$, which completely describe the temperature dependence
of the EFG tensor, Eqs.\ (\ref{m4a})--(\ref{m5}).

%%%%%%%%%%%%%%%%
\subsection{Technical details of calculations}
\label{sub_lapw}

Electron density functional calculations have been performed with the Moscow-FLAPW code \cite{lapw}.
The code explicitly takes into account the nuclear size and the change of the potential and the wave function inside
the nuclear region to obtain the electric field gradient accurately. In addition, the number of radial points
inside the MT-region has been increased to 3000 (for some runs 3500).
The typical LAPW basis set cut-off parameter was $R_{MT} K_{max}$ from 7.8 to 8
with the number of basis functions from 583 to 737 for various runs,
a typical number of $k$ points in the irreducible part of the Brillouin zone (BZ) was 840.
We have used the tetrahedron method for the linear interpolation of energy between $k$ points.
For calculation of the exchange-correlation
potential we have used the PBEsol variant \cite{PBEsol} of the generalized-gradient approximation (GGA) within
the density functional theory (DFT).
The precise calculations of $V_{zz}$ and $\eta$ have been performed for a number of various lattice constants
whose $c/a$ ratios are shown in
Figs.\ \ref{fig3} and \ref{fig4}. The correct behavior of $\eta$ with $T$ has been obtained
for the lattice constants $a=4.5685$ {\AA}, $c=2.9115$ {\AA}, $u = 0.30414$ with the MT-radii 0.8731~{\AA}.
It was assumed that the semicore (Ti $3p$) states followed adiabatically the nuclei vibrations
and their contribution to the interstitial region and temperature dependence was not considered.

For the phonon part required for mean-square displacement calculations, we have used
the pseudopotential method as implemented in Quantum Espresso (QE \cite{QE1,QE2},
with the PBEsol exchange-correlation functional \cite{PBEsol}
and PBEsol optimized norm conserving Vanderbilt pseudopotentials (ONCVPSP) \cite{Ham} from
the ONCVPSP  library \cite{lib1}.
The integration over the BZ for the electron density of states has been performed on a $24 \times 24 \times 12$ grid of $k$-points,
the plane-wave kinetic cut-off energy was 70 Ry.
The lattice-dynamical calculations have been carried out within the density-functional perturbation theory (DFPT).
Phonon dispersions have been computed using the interatomic force constants based on a $6 \times 6 \times 4$ $k$-point grid,
with a $48 \times 48 \times 24$ grid used to obtain the phonon densities of states and
the mean square displacements of titanium and oxygen.

In both cases the thermal expansion of the unit cell has not been taken into account.

%%%%%%%%%%%%%%%%%%%%%%%%%%%%%%%%%%%%%%%%%%%%%%%%%%%%%%%%%%%%%%%%%%%%
\section{Results and discussion}
\label{sec:res}

First, we note that both the expansion of the lattice and a slight increase of $c/a$ with temperature (with the rate 2.2$\times 10^{-6}$~K$^{-1}$ \cite{Sugi})
lead to a smooth decrease of the electric field gradient $V_{zz}=V_{33}$, Fig.\ \ref{fig4} (around $c/a = 0.64$).
Thus, the effect of the increase of EFG observed experimentally \cite{Kaner} is highly nontrivial.

\begin{figure}[!]
\resizebox{0.43\textwidth}{!}
{\includegraphics{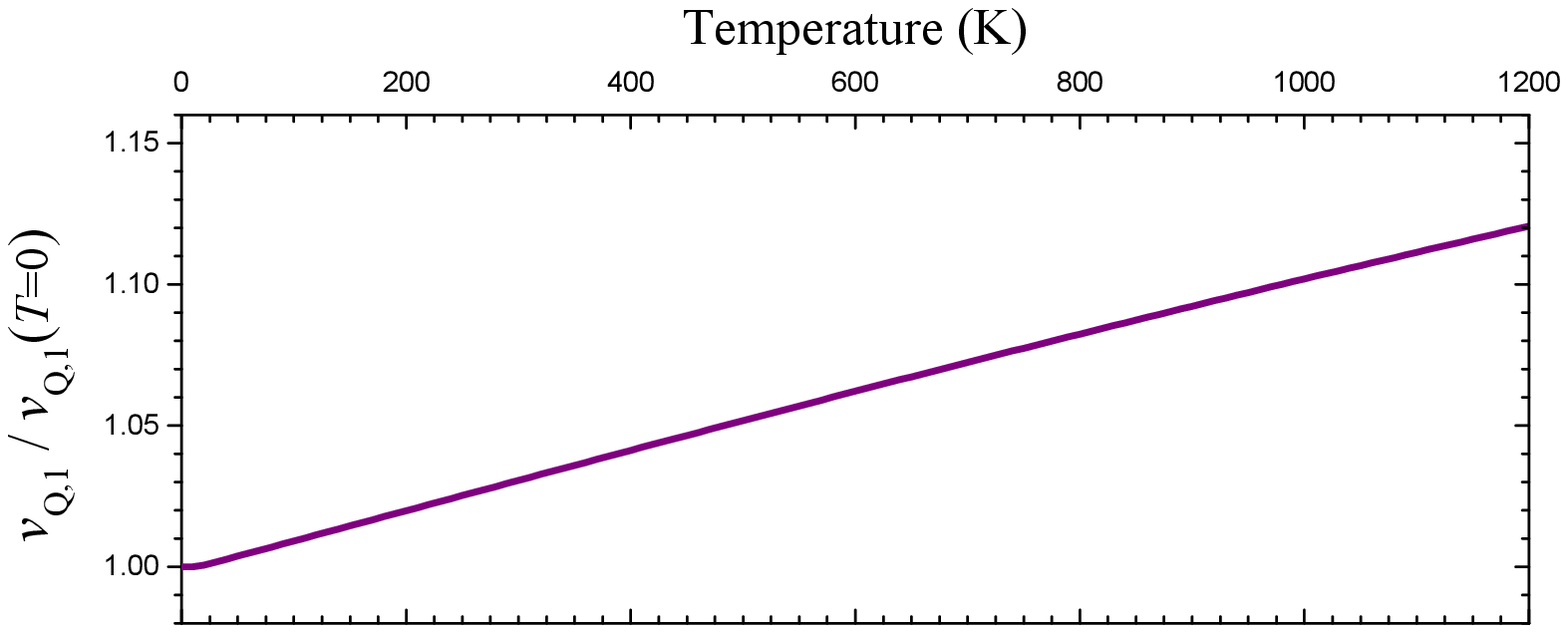}}
\vspace{2mm}
\resizebox{0.43\textwidth}{!}
{\includegraphics{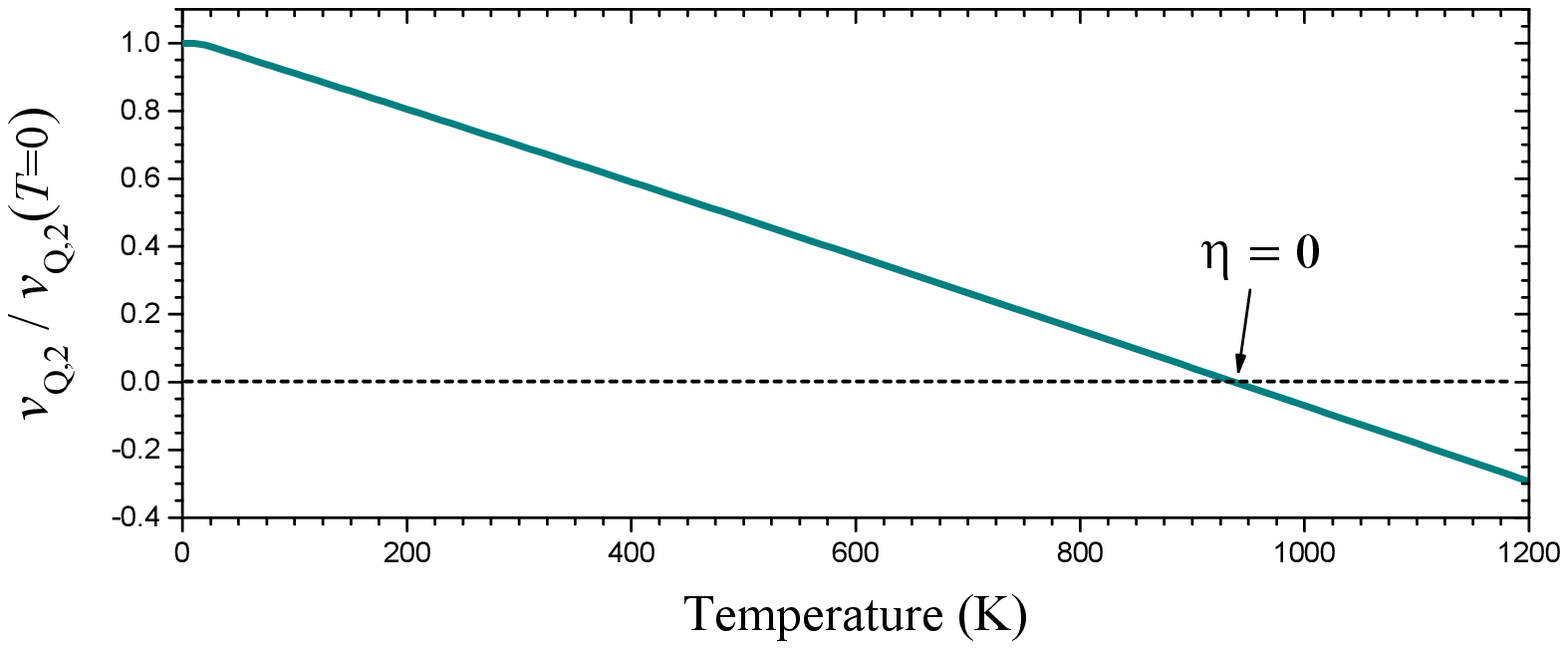}}

% -----------> Figure Caption
\vspace{2mm}
\caption{
A typical temperature dependencies of quadrupole parameters $v_{Q,1}(T) / v_{Q,1}(0)$ and $v_{Q,2}(T) / v_{Q,2}(0)$, Eq.\ (\ref{m3b}),
related to the quadrupole functions $S_{Q,1}$ and $S_{Q,2}$, Eqs.\ (\ref{e1a}) and (\ref{e1b}).
} \label{fig5}
\end{figure}

Our typical calculation of the temperature evolutions of the potential quadrupole quantities $v_{Q,1}$ and $v_{Q,2}$, Eq.\ (\ref{e3}),
are shown in Fig.\ \ref{fig5} (for $c/a=0.6373$).
The increase of $v_{Q,1}$ with $T$ unambiguously results in increase of EFG, Eqs.\ (\ref{m4c}) and (\ref{m5}).
Interestingly, $|v_{Q,2}|$ decreases with $T$ and crosses the zero value at 940 K, after which the sign of $v_{Q,2}$ is reversed.
From Eqs.\ (\ref{m4a}) and (\ref{m4b}) then it follows that at 940 K $\eta = 0$, in close correspondence with experimental
observations \cite{Kaner}.
%%%%%%%%%%%%%%%%%%%%%%%%%%%%%%%%%%%%%%%%%%%%%%%%%%%%%%%%%%%%%%%%%%
%
%------------------------------------------------------
%    FIGURE 5
%------------------------------------------------------
%
%%%%%%%%%%%%%%%%%%%%%%%%%%%%%%%%%%%%%%%%%%%%%%%%%%%%%%%%%%%%%%%%%

The increase of EFG in TiO$_2$ is a very stable effect. It is found in a very wide range of lattice constants and $c/a$ ratios.
The temperature evolution of $v_{Q,2}$ on the other hand is very sensitive to details of calculations.
We have found that at $c/a < 0.641$  $|v_{Q,2}|$ decreases as shown in the lower panel of Fig.\ \ref{fig5},
but increases at $c/a > 0.641$, Fig.\ \ref{fig3}.

The principal mechanism of an increase of EFG has been discussed in Ref.\ \cite{Nik}. Below we briefly discuss it for TiO$_2$.
From Eqs.\ (\ref{m5}) and (\ref{m3}) it follows that $V_{zz}(T) \propto \langle \rho_{Q,1}(R_{MT},T) \rangle$, while
the Fourier expansion for $\langle \rho_{Q,1}(R_{MT},T) \rangle$, Eq.\ (\ref{m1}), being positive, can be further partitioned in two groups,
\begin{eqnarray}
 \langle \rho_{Q,1} \rangle = \sum_m |c_{1,m}|\, {\cal W}_{1,m}(T) - \sum_n |c_{2,n}|\, {\cal W}_{2,n}(T) .
 \label{r1}
\end{eqnarray}
Here ${\cal W}_{i,p}(T)=\exp(-W(\vec{K}_{i,p},\, T))$ ($i=1$, $p=m$ or $i=2$, $p=n$) are the temperature dependent SRDW factors, while
$c_{i,p}(K_{i,p})=-4\pi j_2(K_{i,p}R_{MT})S_{Q,i}(\hat{K}_{i,p})\, \rho(\vec{K}_{i,p})$ are temperature independent coefficients,
which can be positive or negative.
[$\rho(\vec{K}_{i,p})$ here are the Fourier coefficients entering also Eq.\ (\ref{m1}).]
The difference between two explicitly written groups in Eq.\ (\ref{r1}) is that for all indices $m$ and $n$, $c_{1,m} \geq 0$
whereas $c_{2,n} < 0$, and therefore, each of the two sums on the right hand side of Eq.\ (\ref{r1}) is positive.
While SRDWFs ${\cal W}_{i}$ and consequently the both sums always reduce with $T$, the whole sum for TiO$_2$ increases.
This occurs because the second sum in Eq.\ (\ref{r1}) drops fast enough to overcompensate the reduction of the first.

A detailed description of an increase of EFG is complicated: it clearly depends on the lattice structure
through reciprocal vectors $\vec{K}$ and connected with them numerical quantities $c_{i,p}$ (expressed in $j_2(K)$, $S_{Q,i}(\hat{K})$),
and also on the Fourier density components $\rho(\vec{K})$. All these quantities have been calculated numerically
with the Moscow-FLAPW code \cite{lapw}, details are given in Sec.\ \ref{sub_lapw}.
Other important quantities giving the explicit temperature dependence in SRDWFs are mean square displacements $\langle u^2 \rangle$ of titanium, Eq.\ (\ref{m2}).
In the case of the rutile phase of TiO$_2$ there are three independent quantities: $U_{11}=\langle u_x^2 \rangle$, $U_{12}=\langle u_x u_y \rangle$,
and $U_{33}=\langle u_z^2 \rangle$, since by symmetry $U_{22}=U_{11}$ and $U_{13}=U_{23}=0$. Then $W(\vec{K}, T)$, Eq.\ (\ref{m2}), reads as
\begin{subequations}
\begin{eqnarray}
   W(\vec{K}, T) = \frac{1}{2} \left( (K_x^2 + K_y^2)\, U_{11}(T) + 2 K_x K_y\, U_{12} \right. \nonumber \\
    \left. + K_z^2\, U_{33}(T)  \right) .
  \label{r2a}
\end{eqnarray}
It can be brought in the diagonal form,
\begin{eqnarray}
   W(\vec{K}, T) = \frac{1}{2} \left( K_{x'}^2\, U_{1'1'}(T) + K_{y'}^2\, U_{2'2'}(T) + K_z^2\, U_{33}(T)  \right) . \nonumber \\
  \label{r2b}
\end{eqnarray}
\end{subequations}
Here $K_{x'}$ and $K_{y'}$ are projections of $\vec{K}$ on new diagonal crystallographic axes $[1\bar{1}1]$ and $[111]$, while the
third axis ($z'=z$) remains unchanged, and $U_{1'1'}= U_{11} - U_{12}$, $U_{2'2'}= U_{11} + U_{12}$.
Notice, that due to our initial choice of crystallographic axes, Fig.\ \ref{fig1}, the diagonal axes $x'$ and $y'$ are interchanged
in comparison with those in experimental Refs.\ \cite{Burd,Hill,How,Gon,GF,AB,Shin} and consequently $U_{12} > 0$.

\begin{figure}[!]
\resizebox{0.45\textwidth}{!}
{\includegraphics{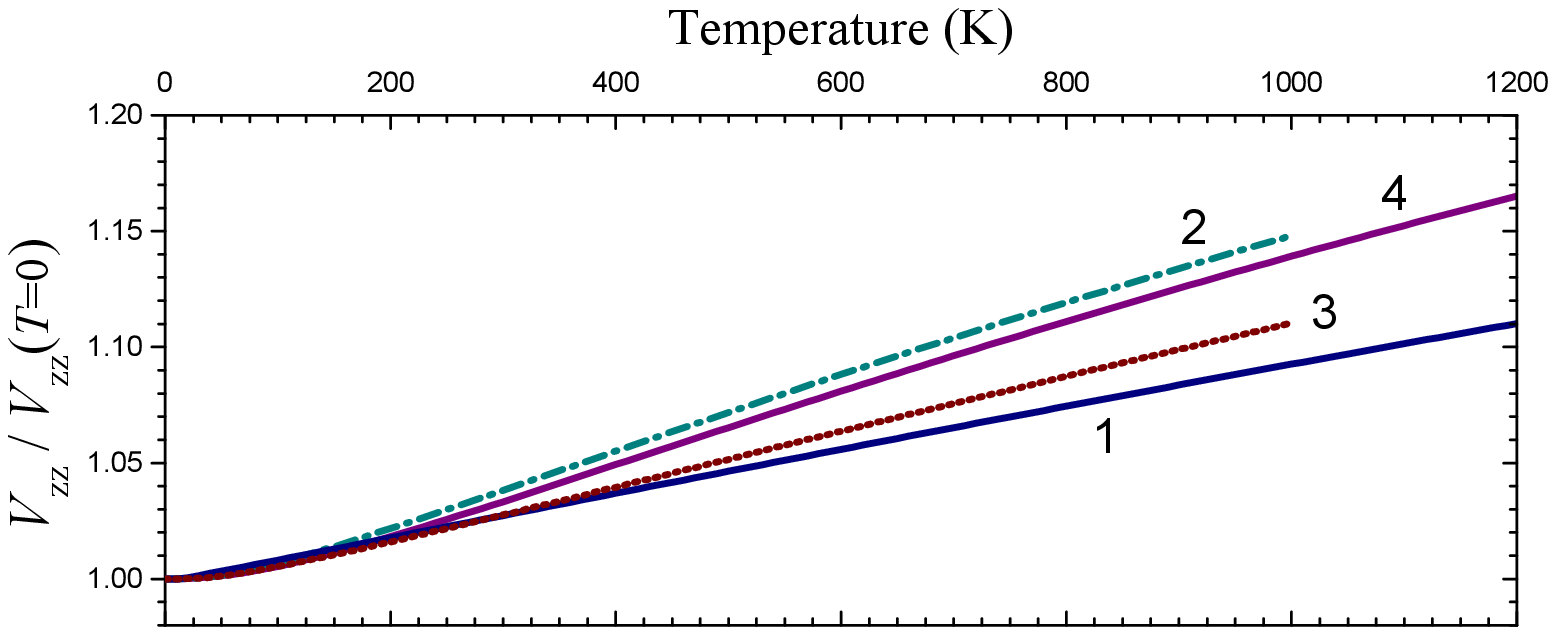}}
%\vspace{2mm}
\resizebox{0.45\textwidth}{!}
{\includegraphics{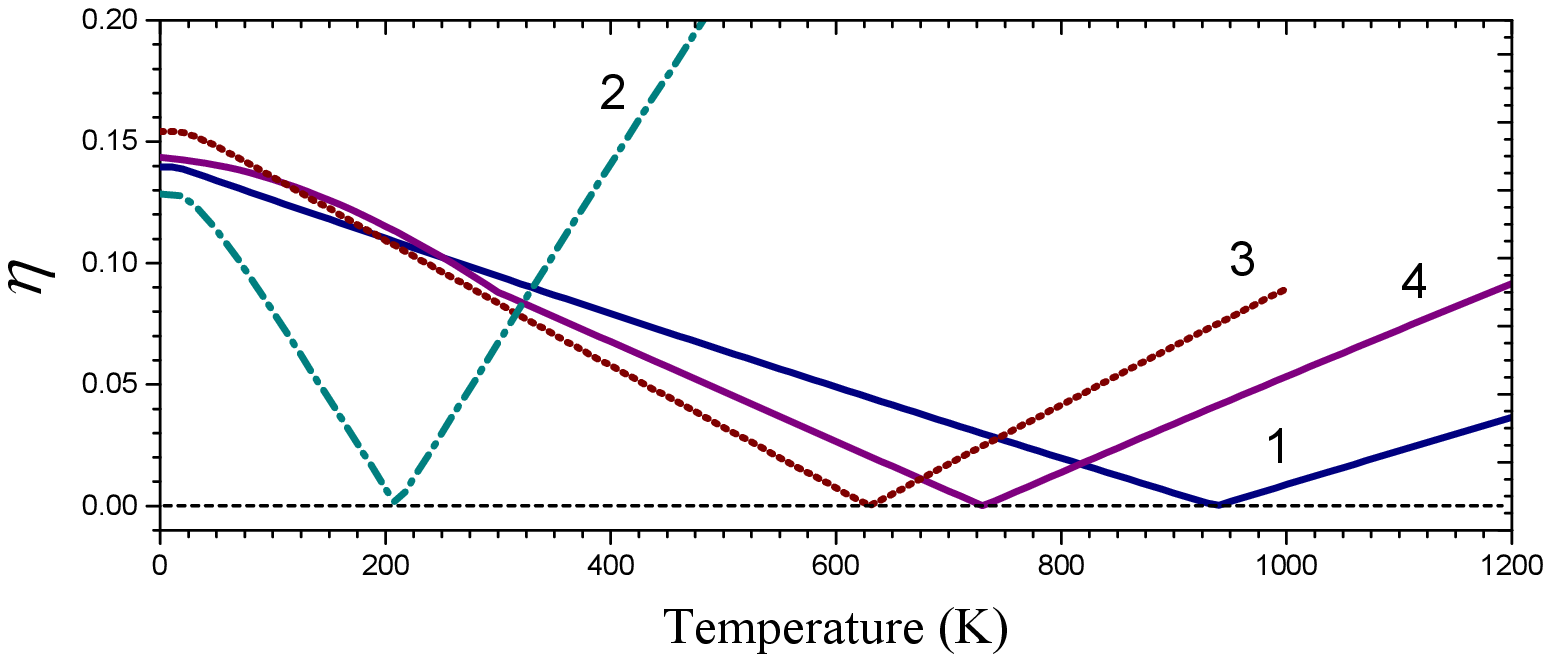}}

% -----------> Figure Caption
%\vspace{2mm}
\caption{
Temperature evolution of $V_{zz}$ and $\eta$ calculated within three models for mean square displacements ($U_{11}$, $U_{12}$, $U_{33}$).
Model 1 -- the linear approximation for MSDs with the values at $T=15$ and $T=295$~K taken from Ref.\ \cite{Burd}.
Model 2 -- MSDs from the QE calculations (harmonic approximation).
Model 3 -- the scaled MSDs from the QE calculations.
Model 4 -- the modified Debye approximation ($T_D$=600~K \cite{How}) Calculation with $c/a=$0.6373.
} \label{fig6}
\end{figure}

A precise calculation of the mean square displacements $U_{11}$, $U_{12}$ and $U_{33}$ (or alternatively, $U_{1'1'}$, $U_{2'2'}$, and $U_{33}$)
is a formidable problem. Even experimental data obtained from fitting neutron and x-ray diffraction study of the rutile using both single crystal
and powder samples of TiO$_2$ at room temperature are quite different, Table \ref{tab1}.
%%%%%%%%%%%%%%%%%%%%%%%%%%%%%%%%%%%%%%%%%%%%%%%%%%%%%%%%%%%%%%
%
%   TABLE 1
%   -------
\begin{table}
\caption{Experimental mean-square displacements (in units $10^{-3} \times${\AA}$^2$) of titanium in TiO$_2$.
$U_{1'1'}$, $U_{2'2'}$ and $U_{3'3'}= U_{33}$ are
the diagonal values of the Ti thermal ellipsoid in principal axes. r.t. stands for the room temperature.
$U_{12} > 0$ in correspondence with the chosen coordinate axes, Fig.\ \ref{fig1}. \\
$^*$ MSDs converted from $\beta_{11}$, $\beta_{33}$, $\beta_{12}$.
\label{tab1} }

\begin{ruledtabular}
\begin{tabular}{l | c  c  c  c  c }

  Ref., $T$ & $U_{11}$  &  $U_{12}$   &  $U_{1'1'}$  & $U_{2'2'}$ & $U_{33}$ \\
\tableline
 \cite{Burd}, 15 K  & 1.2  &  0.2  & 1.0  & 1.4  & 1.1 \\
 \cite{Burd}, 295 K & 5.5  &  0.3  & 5.2  & 5.8  & 4.5 \\
 \cite{How}, r.t.   & 6.8  &  0.4  & 6.4  & 7.2  & 4.6 \\
 \cite{Gon}, r.t.   & 6.68 &  0.12 & 6.56 & 6.80 & 4.81 \\
 \cite{GF}, r.t.    & 6.9  & 0.0   & 6.9  & 6.9  & 4.4 \\
 \cite{AB}$^*$, 298 K   & 6.41 & 0.11  & 6.30 & 6.52 & 4.04 \\
 \cite{Shin}, 299 K & 6.99 &  0.31 & 6.68 & 7.30 & 4.67 \\
 \cite{Sugi}$^*$, 298 K & 7.64 &  0.06 & 7.58 & 7.70 & 4.21 \\

\end{tabular}
\end{ruledtabular}
\end{table}
%%%%%%%%%%%%%%%%%%%%%%%%%%%%%%%%%%%%%%%%%%%%%%%%%%%%%%%%%%%%
Almost all of these studies refer to the room temperature values of MSDs.
To the best of our knowledge in the literature there is only one low temperature (15~K) neutron powder diffraction investigation of TiO$_2$
carried out by Burdett et al. \cite{Burd}, whose data are also quoted in Table \ref{tab1}.
With these two point MSD data taken at 15~K and 295~K \cite{Burd} assuming the linear dependence in all temperature range, one can construct the temperature dependence
of $U_{11}(T)$, $U_{12}(T)$ and $U_{33}(T)$. We shall refer to this $T$-dependence of MSDs as model 1. The EFG parameters obtained with these MSDs
are depicted in Fig.\ \ref{fig6}.

We have also performed the DF perturbation calculation of lattice dynamics with the QE package, details of which are described in Sec.~\ref{sub_lapw}.
The results are shown in Fig.\ \ref{fig6}, model 2.
In units $10^{-3} \times${\AA}$^2$ for 15~K we have obtained $U_{11}$=2.62, $U_{12}$=0.06, $U_{33}$=1.85, and for 300~K $U_{11}$=12.85, $U_{12}$=0.52, $U_{33}$=6.89.
Thus, the calculated values of MSD are significantly overestimated in comparison with experimental ones, Table \ref{tab1}. 
There are several reasons for this discrepancy.
First, the harmonic calculation underestimates the experimental phonon energies \cite{Wehi}. 
In particular, the discrepancy reaches to nearly 5 meV for the low lying transverse acoustic (TA) branch
along the $\Gamma-Z$ and $M-Z$ directions \cite{Wehi}. 
Second, for the TA branch almost linear increase of phonon energies with temperature has been detected \cite{Wehi}
which implies a stiffening of the entire TA branch with $T$ \cite{Lan}.  
In addition, in Refs.\ \cite{Lan,Wehi} a strong role of anharmonicity has been found 
with the phonon mode potential having strong positive fourth order components, which leads to an increase
of the phonon frequency for large atomic displacements occurring at higher temperatures. 
All these effects should result in a substantial decrease of MSDs calculated within the harmonic approximation at room temperature and above. 
% The numerical problem of QE is connected with the fact that TiO$_2$ is polar crystal where the Born effective charge contribution to the dynamical matrix is very large.
% Experimentally, this difficulty is caused by a soft transverse-optical mode in TiO$_2$ \cite{Rest}.

For a more realistic representation of the temperature evolution of MSD, we have scaled our QE data. The scaling factor has been chosen
to fit the room temperature MSD values of Ref.\ \cite{Gon}, Table \ref{tab1}. We refer to these $T$-dependence of MSDs as model 3, Fig.\ \ref{fig6}.
%%%%%%%%%%%%%%%%%%%%%%%%%%%%%%%%%%%%%%%%%%%%%%%%%%%%%%%%%%%%%%%%%%
%
%------------------------------------------------------
%    FIGURE 6
%------------------------------------------------------
%
%%%%%%%%%%%%%%%%%%%%%%%%%%%%%%%%%%%%%%%%%%%%%%%%%%%%%%%%%%%%%%%%%

Finally, we can obtain the temperature dependence of MSDs using a modified Debye model. In Ref.\ \cite{How}
it has been shown that the model of isotropic vibrations of average amplitude with the Debye temperature $T_D$=600~K \cite{How} fits the diffraction data.
Within the Debye approximation we can calculate the average value of MSDs ($U_{av}=(2 U_{11} + U_{33})/3$).
We get $U_{av} = 2.72 \cdot 10^{-3}$ {\AA}$^2$ at 15~K and $U_{av} = 6.00 \cdot 10^{-3}$ {\AA}$^2$ at 300~K.
Based on the $T$-dependence of $U_{av}$ and using the linear interpolation for the ratios $U_{11} / U_{33}$ and $U_{12} / U_{33}$ in the range
from 15~K to 300~K with the data of Ref.\ \cite{Burd}, we can obtain the anisotropic values of MSDs ($U_{11}$, $U_{12}$, $U_{33}$)
at all temperatures. This is the modified Debye model marked as model 4 in Fig.\ \ref{fig6}.

\begin{figure}[!]
\resizebox{0.45\textwidth}{!}
{\includegraphics{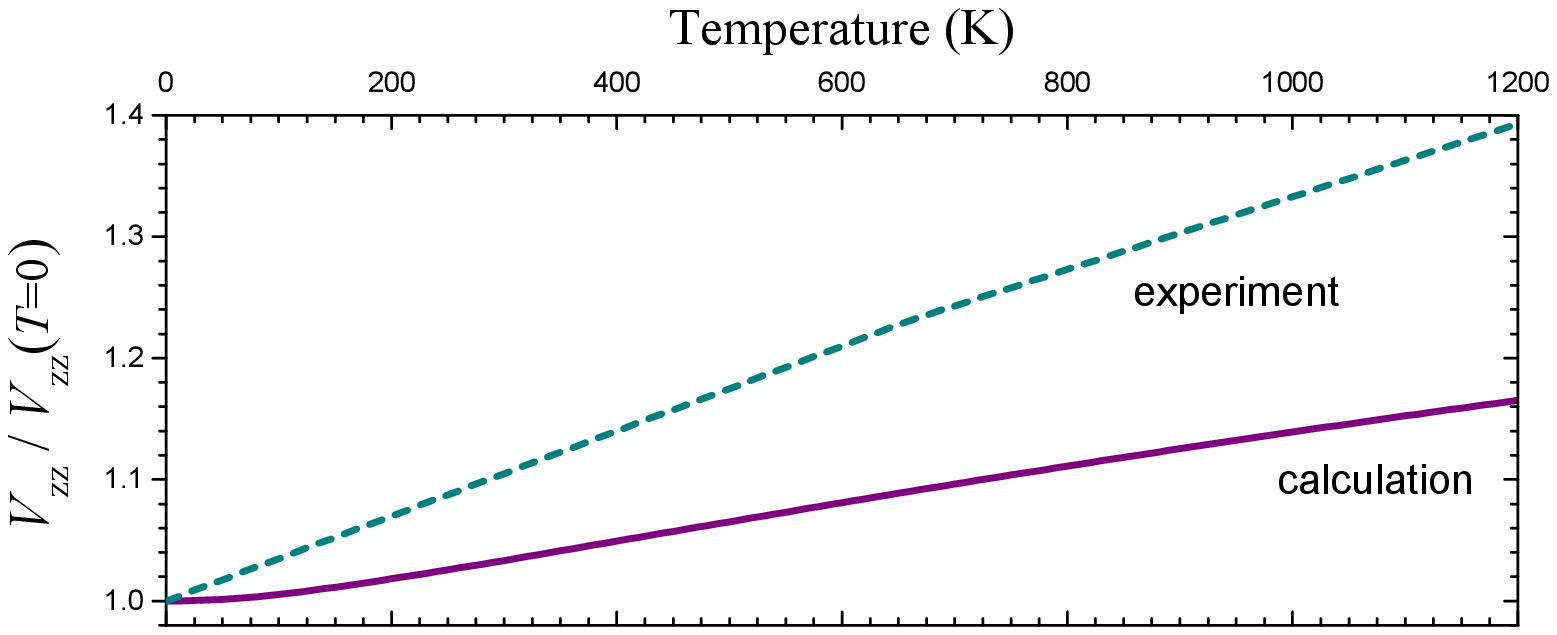}}
\vspace{2mm}
\resizebox{0.45\textwidth}{!}
{\includegraphics{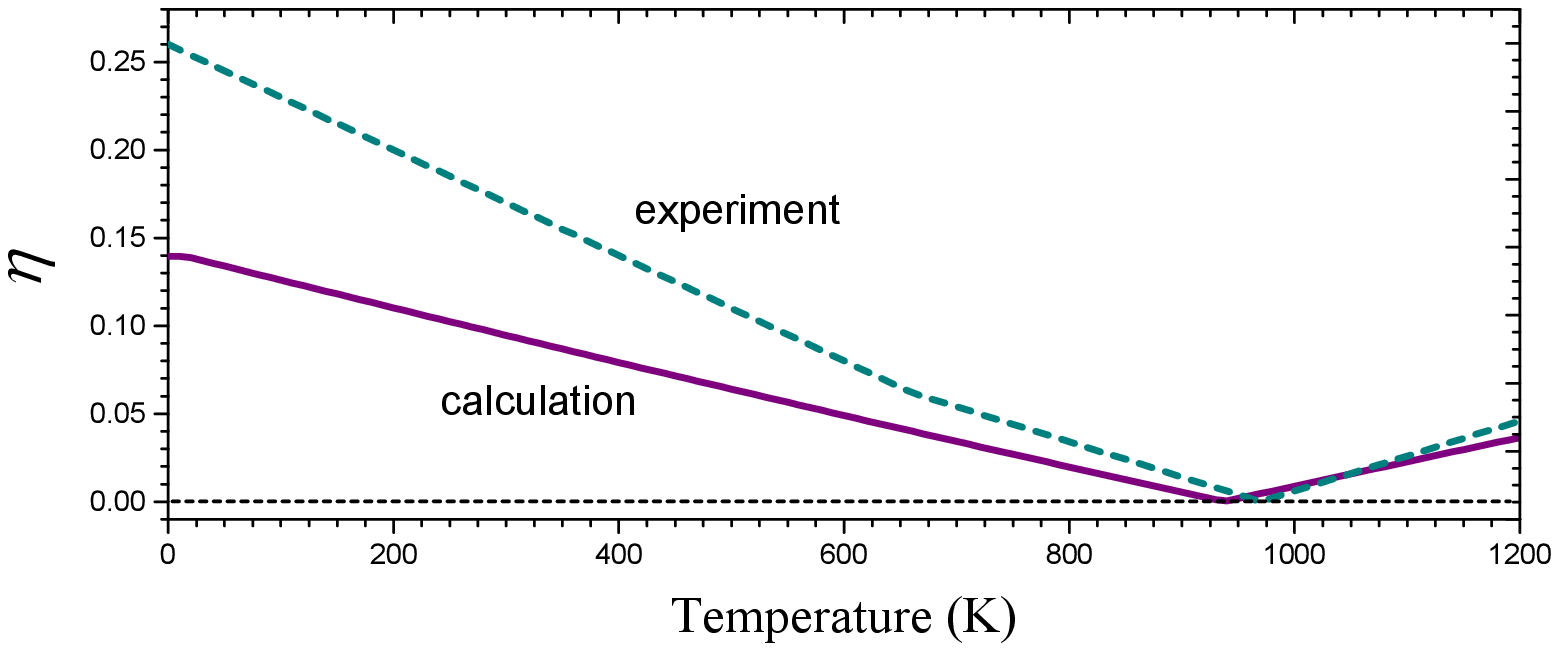}}

% -----------> Figure Caption
%\vspace{2mm}
\caption{
Comparison with the NMR experimental data, Ref.\ \cite{Kaner}.
The upper panel corresponds to the model 4, the lower panel - to the model 1.
} \label{fig7}
\end{figure}

All four models for mean square displacements result in a growth of EFG with temperature and a characteristic dependence
of the asymmetry parameter $\eta$, Fig.\ \ref{fig6}.
Numerically, the plots vary, reflecting a difference in the $T$-evolution of mean square displacements.
In Fig.\ \ref{fig7}, the best calculated dependencies for EFG and $\eta$ are compared with NMR data of Kanert and Kolem \cite{Kaner}.
%%%%%%%%%%%%%%%%%%%%%%%%%%%%%%%%%%%%%%%%%%%%%%%%%%%%%%%%%%%%%%%%%%
%
%------------------------------------------------------
%    FIGURE 7
%------------------------------------------------------
%
%%%%%%%%%%%%%%%%%%%%%%%%%%%%%%%%%%%%%%%%%%%%%%%%%%%%%%%%%%%%%%%%%

The theoretical increase of $V_{zz}$ is approximately half of
the experimental one. We ascribe it to anharmonic effects which were found important
for the description of the temperature evolution of EFG in Zn and~Cd~\cite{Nik}. 
EFG there is found to be very sensitive to the correct temperature evolution of the ratio $U_{33}(T) / U_{11}(T)$,
which could be the case for TiO$_2$ taking into account a stiffening of its low lying phonon energy spectrum with $T$ 
and a strong forth order component of the potential \cite{Lan,Wehi}.
Another important factor is a possible change of the oxygen rutile position $u$ with $T$,
which greatly influence the value of EFG \cite{Ryu}.

The calculated $V_{zz}$ is in good correspondence with the experimental value of 2.0$\times 10^{21}$~V/m$^2$,
obtained by extrapolating experimental data to $T=0$ \cite{Kaner}.
The zero-point vibrations lead to a 0.5 {\%} increase of the calculated EFG,
to the corrected value 1.998 $\times 10^{21}$~V/m$^2$ at the experimental ratio $c/a=0.642$.

%%%%%%%%%%%%%%%%%%%%%%%%%%%%%%%%%%%%%%%%%%%%%%%%%%%%%%%%%%%%%%%%%%%%
\section{Concluions}
\label{sec:conc}

On the basis of {\it ab initio} calculations of the rutile phase of TiO$_2$ we have reproduced qualitatively the temperature
increase of the electric field gradient $V_{zz}$ and the evolution of the asymmetry parameter $\eta$, Fig.\ \ref{fig7}.
Note that if only the anisotropic thermal expansion of the lattice parameters is taken into account, i.e. an increase
of $c/a$, this would predict a decrease of $V_{zz}$.
Our consideration relies on the method used in Ref.\ \cite{Nik} for calculation of the temperature
dependence of EFG in Zn and Cd.
The method utilizes the effect of changing the averaged value of the quadrupole components $\langle \rho_{Q,1} \rangle$
and $\langle \rho_{Q,2} \rangle$, Eq.\ (\ref{m1}), on a titanium MT-sphere vibrating with the Ti nucleus.
[The angular quadrupole functions $S_{Q,1}$ and $S_{Q,2}$ are explicitly given by Eqs.\ (\ref{e1a}) and (\ref{e1b}).]
The temperature dependence is introduced through the factor $\exp(-W(\vec{K},T))$
which is the square root of the usual Debye-Waller factor \cite{Bru}, whereas
the thermal expansion of the unit cell has not been taken into account.
SRDWF effectively reduces the corresponding Fourier component of the electron density, Eq.\ (\ref{m1}),
and depends sensitively on the independent mean square displacements $U_{11}(T)$, $U_{12}(T)$ and $U_{33}(T)$.

$U_{11}(T)$, $U_{12}(T)$ and $U_{33}(T)$ have been found experimentally, Table \ref{tab1},
in neutron and x-ray diffraction studies at room temperature and at 15~K \cite{Burd}.
In addition, we have computed MSDs with Quantum Espresso in the temperature range from zero to 1000 K, but
these values of $U_{11}(T)$ and $U_{33}(T)$ have turned out to be overestimated in comparison
with the experimental ones.
On the basis of experimental and calculated MSD data, we have adopted four models for their $T$-dependence,
and calculated corresponding $V_{zz}(T)$ and $\eta(T)$, Fig.\ \ref{fig6}.

For all cases our results indicate a stable increase of EFG with temperature, Figs.\ \ref{fig6} and \ref{fig7}.
The increase of EFG is a remarkable effect arising from interplay of positive and negative terms, Eq.\ (\ref{r1}).
While each term is reduced with temperature due to the $\exp(-W(\vec{K},T))$ factors, the whole sum
can growth with $T$, leading to the increase of EFG.
Quantitatively however the calculated growth of $V_{zz}$ is smaller than the experimental one.
This can be explained by an appreciable change of oxygen polarizability with $T$ \cite{Ryu} or
by anharmonic effects playing an important role in Zn and Cd \cite{Nik}.
Such effects include a nonlinear change of the $U_{11}/U_{33}$ and $U_{12}/U_{33}$ ratios with temperature,
a change of oxygen position
and the thermal expansion of the rutile lattice. 
MSDs in TiO$_2$ are apparently influenced by several unusual anharmonic 
effects \cite{Lan,Wehi,Rest} found in this compound. 

For calculations with $c/a < 0.641$ when $V_{22} > V_{11}$, Fig.\ \ref{fig4},
the calculated temperature evolution of $\eta$ shows a good correspondence with
experimental data, Figs.\ \ref{fig7} and \ref{fig6}.
In particular it fairly well reproduces the crossing of the zero value,
which is a nontrivial feature of the experimental curve \cite{Kaner}.
This property is connected with the quadrupole potential parameter $v_{Q,2}$
going through zero at this temperature, Fig.\ \ref{fig5}.
For the $c/a > 0.641$ case corresponding to the experimental $c/a$ ratio,
$\eta$ demonstrates just a monotonic linear increase with $T$
in disagreement with the experiment.

Interestingly, when $c/a > 0.653$, the electric field gradient
switches to another component, $V_{zz}=V_{11}$ in Fig.\ \ref{fig4}.
Note that such a situation is experimentally observed for Cd probes in TiO$_2$
where it is accounted for by local displacements caused by Cd
replacing a Ti atom in the rutile lattice \cite{Erri02,Erri03}.
Although the electronic configuration of Ti and Cd are quite different, the
change in the orientation of $V_{zz}$ from (001) to (110) is produced
by a similar distortion of the nearest neighbor oxygen octahedron.

%%%%%%%%%%%%%%%%%%%%%%%%%%%%%%%%%%%%%%%%%%%%%%%%%%%%%%%%%%%%%%%%%%%%%%%%%%%%
\acknowledgements

This research was supported by a grant of the Russian Science
Foundation (Grant No 18-12-00438). We are grateful to the Polish representative
at the Joint Institute for Nuclear Research.
The numerical computations have been carried out using computing resources of
the federal collective usage center Complex for Simulation and Data Processing
for Mega-science Facilities at NRC “Kurchatov Institute”,
http://ckp.nrcki.ru/,
and supercomputers at Joint Supercomputer Center of RAS (JSCC RAS).
We also thank the Uran supercomputer of IMM UB RAS for access.

%\appendix

%%%%%%%%%%%%%%%%%%%%%%%%%%%%%%%%%%%%%%%%%%%%%%%%%%%%%%%%%%%%%%%%

%---------------- REFERENCES -------------------------------

\end{document}